\documentclass[aps,prb,reprint,groupedaddress,showpacs]{revtex4-1}

\usepackage[colorlinks=true,linkcolor=blue,citecolor=blue,urlcolor=blue]{hyperref}
\usepackage{amsmath}
\usepackage{bbold}
\usepackage{epsfig}
\usepackage{graphics}
\usepackage[section]{placeins}
\usepackage{epstopdf}
\usepackage[english]{babel}
\usepackage{color}
\usepackage{footmisc}

\newcommand*{\citen}[1]{%
  \begingroup
    \romannumeral-`\x 
    \setcitestyle{numbers}%
    \cite{#1}%
  \endgroup   
}

\begin{document}
\title{Resonators coupled to voltage-biased Josephson junctions: \\
From linear response to strongly driven nonlinear oscillations}

\author{S. Meister}
\author{M. Mecklenburg}
\author{V. Gramich}
\altaffiliation{Present address: Fraunhofer-Institut f\"ur Angewandte Fest\-k\"orperphysik, Tullastra{\ss}e 72, 79108 Freiburg, Germany}
\author{J. T. Stockburger}
\author{J. Ankerhold}
\author{B. Kubala}
\affiliation{Institut f\"ur komplexe Quantensysteme and IQST, Universit\"at Ulm,
Albert-Einstein-Allee 11, 89069 Ulm, Germany}

\date{\today}

\begin{abstract}
Motivated by recent experiments, where a voltage biased Josephson junction is placed in series with a resonator, the classical dynamics of the circuit is studied in various domains of parameter space. This problem can be mapped onto the dissipative motion of a single degree of freedom in a nonlinear time-dependent potential, where in contrast to conventional settings the nonlinearity appears in the driving while the static potential is purely harmonic. For long times the system approaches steady states which are analyzed in the underdamped regime over the full range of driving parameters including the fundamental resonance as well as higher and sub-harmonics. Observables such as the dc-Josephson current and the radiated microwave power give direct information about the underlying dynamics covering phenomena as bifurcations, irregular motion, up- and down conversion. Due to their tunability, present and future set-ups provide versatile platforms to explore the changeover from linear response to strongly nonlinear behavior in driven dissipative systems under well defined conditions.
\end{abstract}


\maketitle

\section{Introduction\label{sec:Introduction}}
The nonlinear properties of Josephson junctions (JJs) have made such devices a key circuit element for classical and quantum electronics. Accordingly, there has been a long tradition of studying non-linear phenomena in driven superconducting circuits, starting as early as the 1960s with the discovery of Shapiro steps \cite{Barone82}. While Shapiro-steps have remained a tool in exploring new directions in Josephson physics, for instance in atomic point contacts \cite{Goffman00,Cuevas02,Chauvin06,Bergeret11,Bretheau12}, other nonlinear phenomena, like synchronization, have been investigated in arrays of JJs: as a test-bed for generic theory models to capture synchronization phenomena \cite{Wiesenfeld96, Wiesenfeld98}, but as importantly with the prospect of applications as sources of more intense coherent radiation \cite{Barbara99}, cf. also new developments using intrinsic arrays \cite{Ozyuzer07, Wang09, Welp13}.

More recently, the nonlinearity of the JJ was exploited as the crucial factor in enabling the high sensitivity  of Josephson bifurcation amplifiers, achieving substantial improvements towards reaching   quantum-limited measurement processes \cite{Siddiqi04,Siddiqi05,Vijay09,Boulant07}. Most of the features of Josephson bifurcation amplifiers, in fact, only rely on (and can consequently be described by) any type of nonlinearity  \cite{Verso10,Peano04, Serban07, Marthaler06, Kohler97, Dykman12}, e.g., Duffing-type models, so that only recently the full nonlinear potential of the JJ has become of interest in this field \cite{Jung14}.

A recent addition to the field of driven nonlinear JJs \cite{Werthamer67,Astafiev07, Chen11} are experiments on a dc-biased JJ connected to a resonator \cite{Hofheinz11,Blencowe12,Chen14}. In this sort of setup, charge transfer through the JJ leads to excitations in the resonator, and therefore allows to convert energy carried by charge quanta into quantum microwave photons. In these devices measurement of both the Josephson current and the emitted microwave radiation is possible, a distinct advantage in comparison to other recently proposed transport setups \cite{Monis12,Hussein10}, which  show similar nonlinear features like bifurcations, period multiplication and up- and down-conversion \cite{Baker96, Guckenheimer93, Iooss97}. Such nonlinear effects will, in fact, dominate the system's dynamics and therewith the characteristics of the Josephson current and the emitted microwave radiation for driving beyond a linear regime of weak Josephson coupling.

While quantum properties of the system, in particular of the emitted microwave radiation have been investigated widely, both theoretically \cite{Padurariu12, Armour13, Gramich13, Leppakangas13, Leppakangas14, Armour15, Kubala15, Dambach15, Trif15, Guo14, Leppakangas15} and within new experiments \cite{Portier15}, due to the complexity and richness of nonlinear effects  a deeper understanding of the purely classical dynamics of the system is instructive, but also of relevance for current experimental activities\citep{Chen14}, as we will discuss below. This is especially the case, as the nonlinearity enters the system in a peculiar way. It does not stem from a nonlinear (static) potential, but rather from the manner of coupling the drive (JJ)  to the resonator. The Josephson phase in this setup is thus not fixed by the external voltage, but appears as a dynamical degree of freedom manifest in a time-dependent effective potential determining the phase dynamics. We note in passing that related nonlinear phenomena have recently been explored for Josephson phase slip devices\cite{Hriscu11}.

In this paper we present analytical and numerical investigations in the regime, where the system's dynamics is described by classical Josephson equations. While the features found are to an extent common to a wide class of nonlinear classical systems, which specific effects in what distinct manner are realized and how they are observed in this new type of nonlinear system is an intriguing open question.
To tackle it, we will first present the system under study in more detail and introduce the analytical methods used in Sec.~\ref{sec:system+formalism}.
 The following sections cover the fundamental resonance (Sec. \ref{sec:Om1}) and higher order resonances (Sec. \ref{sec:Resonances}). The influence of a thermal environment at finite temperatures is investigated in Sec.~\ref{sec:Thermal}, before we conclude in Sec.~\ref{sec:Conclusion}.

\section{Circuit dynamics\label{sec:system+formalism}}
\begin{figure}[t]
	\begin{center}
	\includegraphics[width=8.6cm]{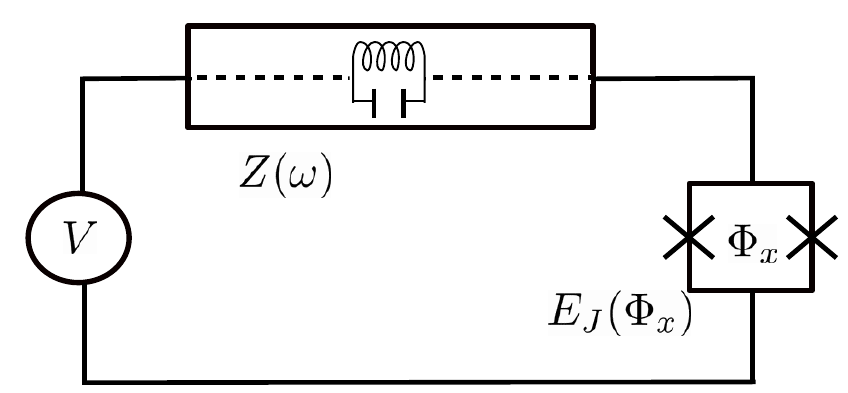}
	\caption{\label{fig:circuit}Circuit diagram of a voltage-biased Josephson junction in series with a resonator with relevant mode frequency $\omega_0$ described by an effective impedance $Z(\omega)$ as specified in (\ref{eq:impediance}). Using a SQUID-geometry\cite{Barone82} for the JJ allows for tuning the effective Josephson energy $E_J(\Phi_x)$ by an external magnetic flux $\Phi_x$.
	}
	\end{center}
\end{figure}

We consider a circuit (see Fig.~\ref{fig:circuit}), where a Josephson junction (JJ) is placed in series with a resonator with only a single mode of frequency $\omega_0=1/\sqrt{LC}$ being relevant. The total impedance $Z_t(\omega)$ seen by the tunneling Cooper pairs consists of the combination of the capacitance $C_J$ of the JJ and the parallel $LC$ resonance with finite $Q$ factor. Since experimentally $C_J\ll C$, one has $Z_t(\omega)\approx Z(\omega)$ with
\begin{equation}\label{eq:impediance}
Z(\omega)=\frac{1}{C} \frac{\omega}{i(\omega^2-\omega_0^2)+\omega\omega_0/Q}\, .
\end{equation}
Based on Kirchhoff rules, equations of motion for the circuit, biased by a dc-voltage $V$, are found,
\begin{equation}\label{eq:eom1}
  \ddot{\varphi} +\omega_0^2 \varphi+ \omega_0 \frac{\dot{\varphi}}{Q}+(E_J/\phi_0^2 C) \sin(\phi)=0\ , \ \phi=\varphi+\omega_J t\, ,
\end{equation}
expressed in terms of the resonator's phase variable $\varphi=-(2e/\hbar) \int dt V_{LC}(t)$, and the Josephson phase $\phi=(2e/\hbar)\int dt V_J(t)$ with $V=V_{LC}+V_J$.\cite{Barone82, Wendin05} Here, $\omega_J=2 eV/\hbar$ denotes the driving frequency, $\phi_0=\hbar/2e$ the reduced flux quantum, and the time derivative $\cdot=d/dt$.

This set of equations can be cast  into an equation of motion for a fictitious particle with an effective mass, $m=\phi_0^2 C$, moving in a harmonic potential and coupled to an external time-periodic, position dependent force, i.e.,
\begin{equation}
m \ddot{\varphi} +m\omega_0^2 {\varphi}+m\frac{\omega_0}{Q}\dot{\varphi}+ E_J \sin(\varphi+\omega_J t)=\bar{\xi}(t)\;.
\end{equation}
To incorporate finite temperature effects, as discussed below, thermal noise $\bar{\xi}$ at temperature $T$ is added.\cite{Risken96} It is related to the resonator damping via the fluctuation-dissipation theorem as $\langle \bar{\xi}(t)\bar{\xi}(t')\rangle_\beta=2 m (\omega_0/Q) k_{\rm B} T \delta(t-t')$ and $\langle \bar{\xi}\rangle_\beta=0$.

To explore the 
dynamics of the above Langevin equation, it is convenient to work with dimensionless units, where times are scaled with $\omega_0$ and energies with $m\omega_0^2$. This then leads to
\begin{equation}\label{eq:Langevindimless}
\ddot{\varphi}+\gamma\dot{\varphi}+\varphi+\lambda\sin(\Omega t+\varphi)={\xi}(t)\, ,
\end{equation}
where the dimensionless friction coefficient $\gamma$ is related to the $Q$-factor of the resonator via $\gamma=1/Q$ and we further introduced the dimensionless driving amplitude $\lambda=E_J/m\omega_0^2$ and driving frequency $\Omega=\omega_J/\omega_0$. Throughout the first part of the paper we concentrate on the limit $T=0$ (${\xi}\equiv\bar{\xi}/m\omega_0^2=0$) and discuss the impact of thermal noise later.

This form of the equation of motion, \eqref{eq:Langevindimless}, is the starting point for studying the dynamics of the voltage-biased circuit (cf.~Fig.~\ref{fig:circuit}) in the rest of this paper. In particular, we are interested in the long-time limit, where the
 balance between the dissipative and the driving part of Eq.\eqref{eq:Langevindimless} has pushed the system into time-periodic steady-state orbits. Considering the energy balance of the resonator obtained from Eq.\eqref{eq:Langevindimless},
 \begin{equation}
 \begin{aligned}\label{eq:dissEnergy}
\frac{d}{dt}\left( \frac{\dot{\varphi}^2}{2}+\frac{\varphi^2}{2}\right)&=-\gamma\dot{\varphi}^2 &&-\lambda \dot{\varphi}\sin(\varphi+\Omega t)\\
&= \;P_\mathrm{diss} &&+ P_{\mathrm{JJ}\rightarrow{HO}}\,,
\end{aligned}
\end{equation}
we easily identify the power dissipated from the resonator, $P_\mathrm{diss}$, and the power injected into the resonator via the driving, $P_{{\rm JJ}\rightarrow {\rm HO}}\equiv I_J V_{LC}$, which will both be considered in detail below.
The dissipated power together with the dimensionless Josephson current $I_J=\lambda\sin(\varphi +\Omega t)$ constitute the main observables, which can be accessed experimentally, either averaged over many oscillation periods, or time- or frequency resolved. Note that the physical current results from the dimensionless current by multiplying $\phi_0/L$.

Due to the nonlinearity present in \eqref{eq:Langevindimless} the structure of steady-state orbits will depend sensitively on the damping and amplitude and frequency of the driving, giving rise to the full wealth of nonlinear phenomena such as bifurcations, up- and down-conversion etc. What distinguishes the situation under consideration here from most other driven nonlinear systems is that the nonlinearity appears not in form of a static potential energy but rather as part of the driving force. In fact, it turns out that the effective time-dependent potential giving rise to (\ref{eq:Langevindimless}), namely,
\begin{equation}\label{eq:poteff}
V_{\rm eff}(\varphi)=\frac{1}{2}\varphi^2 -\lambda \cos(\varphi+\Omega t)\,,
\end{equation}
can sometimes be illuminating  to achieve a better understanding of the fictitious particle dynamics.

Of course, for nonlinear, time-dependent problems such as \eqref{eq:Langevindimless} explicit solutions can, in general, only be obtained numerically. However, analytical progress, at least 
\begin{figure*}[t!]
	\begin{center}
	\includegraphics[width=0.9\textwidth]{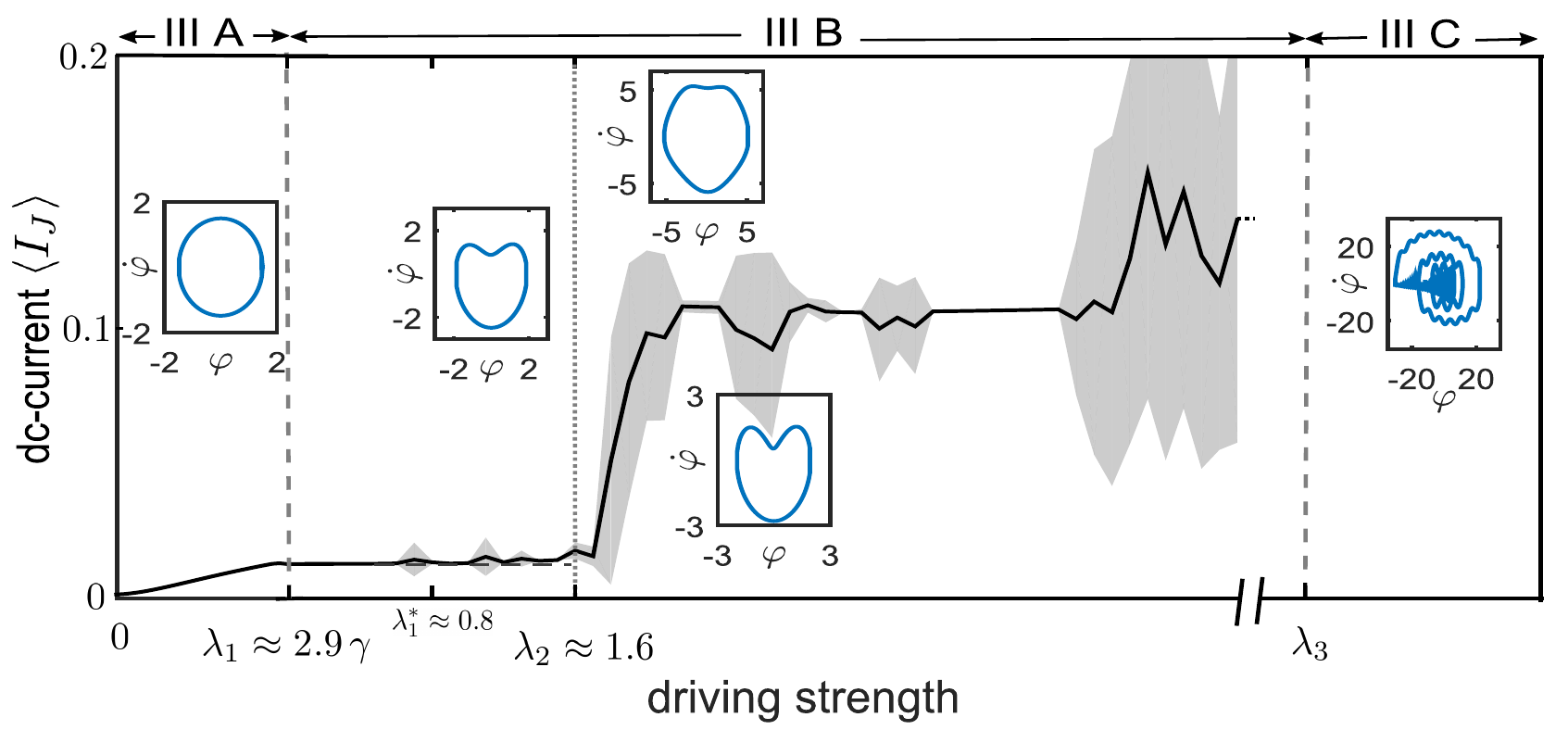}
	\caption{(color online) Tuning the dimensionless Josephson coupling $\lambda$ the steady-state dynamics at the fundamental resonance, $\Omega =\omega_J/\omega_0=1$, accesses different ranges discussed in Secs.~\ref{sec:firstBifurc}, \ref{sec:secondBifurc}, \ref{sec:elevator}, which are characterized by typical phase space portraits.
Within the domain \ref{sec:secondBifurc}, the dotted vertical line separates the range of regular dynamics from the range $\lambda_2<\lambda$, where the transition towards irregular dynamics occurs.
The solid line illustrates the corresponding behavior of the dimensionless dc-current through the JJ with the shadow indicating its standard deviation when starting the steady-state dynamics with  thermal initial conditions. The specific data shown for $\gamma=0.01$ exemplify the generic behavior in the underdamped regime. In this regime, the parameters $\lambda_1/\gamma,\ \lambda_1^*$ and $\lambda_2$ are not very sensitive to varying the friction strength.}
	\label{fig:OverviewSec3}
	\end{center}
\end{figure*}
for the steady state, can be made in limiting domains of parameter space.  Physically, one expects a steady-state solution oscillating with the frequency of the drive, as well as higher and possibly subharmonics. Putting such an ansatz into \eqref{eq:Langevindimless}, the nonlinear sin-term will again produce higher harmonics with the coefficients of the various frequency components appearing in Bessel-functions.

Formally, one can write a generic ansatz,
\begin{equation}\label{eq:ansatz}
\varphi(t)=\varphi_0+\frac{1}{2}\sum_{k\in M_\Omega}^{\hspace{0.6cm}\prime}\varphi_k \, {\rm e}^{ik\Omega t+i\theta_k}\, ,
\end{equation}
where $\varphi_k=\varphi_{-k}$ and $\theta_k=-\theta_{-k}$ so that solutions are real-valued.
The sum runs over a suitable set of rational numbers $M_\Omega$ with the prime indicating that $k=0$ is excluded.
For example, below we will study the fundamental resonance at $\Omega=1$ implying $M_\Omega=\mathbf{Z}$. Other situations are driving at $\Omega=n$ ($n$ integer) with $M_\Omega=\{\nu/n| \nu\in \mathbf{Z}\}$ and at $\Omega=1/n$ ($n$ integer) with $M_\Omega=  \mathbf{Z}$.   Of course, in a perturbative treatment only a finite number of coefficients is taken into account.

Now,  inserting this ansatz into (\ref{eq:Langevindimless}) yields a nonlinear equation for the steady-state Fourier coefficients, i.e.,
\begin{equation}\label{eq:BewglmitAnsatz}
\begin{split}
\varphi_0+\frac{1}{2}\sum\limits_k^{\hspace{0.6cm}\prime} \left[\varphi_k {\rm e}^{i\theta_k}(-\Lambda_k+i\gamma\Omega k)\right] {\rm e}^{ik\Omega t} &\\
+\frac{\lambda}{2i}\left[ {\rm e}^{i\Omega t} {\rm e}^{i\varphi_0}F_+(t)-{\rm e}^{-i\Omega t} {\rm e}^{-i\varphi_0}F_-(t) \right]&=0,
\end{split}
\end{equation}
with $\Lambda_k=k^2\Omega^2-1$ and $F_{\pm}(t)=\prod\limits_{k>0}F_k^\pm(\varphi_k,t)$. These latter functions
\begin{equation}\label{eq:fpmbessel}
F_k^\pm (\varphi_k,t)=\sum_{l=-\infty}^{\infty}i^l(\pm 1)^lJ_l(\varphi_k)e^{il(k\Omega t+\theta_k)}
\end{equation}
contain  Bessel functions $J_l(\cdot)$ of the first kind and of integer order. Eq.~(\ref{eq:BewglmitAnsatz}) serves as a starting point for perturbative treatments in various ranges of the dynamics in order to gain a deeper understanding of the numerical findings based on (\ref{eq:Langevindimless}).

\section{Fundamental resonance \label{sec:Om1}}
We will start in this section with the dynamics near and at the fundamental resonance where the driving frequency  matches the Josephson frequency so that $\Omega= 1$. According to (\ref{eq:Langevindimless}), in absence of noise only two dimensionless parameters are left which determine the nature of steady-state orbits, namely, the friction $\gamma$ and the driving amplitude $\lambda$. Current experimental realizations are operated in the underdamped regime with a fixed $\gamma\ll 1$ ($Q$-factors vary from 10 to about 1000) and varying driving strengths. In the classical domain that we consider here, this leads from simple linear to strongly nonlinear phase space patterns, the structure of which is reflected in specific observables such as charge current and photon flux.

Before we study the details, let us give a brief qualitative account of the different dynamical ranges when $\lambda$ is varied (cf.~Fig.~\ref{fig:OverviewSec3}):\\
(i) In the regime of weak driving (Sec.~\ref{sec:firstBifurc}), the dynamics changes from linear to nonlinear towards a threshold $\lambda_1$, where a first bifurcation occurs.  Perturbative treatments capture this transition fairly accurate.
While the oscillation amplitude already reflects nonlinearities of the system, phase-space orbits in this regime are basically ellipses with only small deviations from the harmonic limit.
Physically, the initial quadratic rise of the dc-current through the JJ flattens with increasing $\lambda$ until it saturates at $\lambda=\lambda_1$. \\
(ii) For an intermediate range $\lambda_1<\lambda<\lambda_2$ (Sec.~\ref{sec:secondBifurc}) the dynamics is dominated by the full nonlinearity of the driving force but remains still regular. Within this domain further bifurcations occur, however, without affecting the dc-current through the JJ which stays basically at its value somewhat above $\lambda_1$. Phase-space orbits increasingly loose their harmonic-like ellipsoidal structure and become potato-shaped with deformations in form of `dips', thus reflecting the fact that the effective time-dependent potential $V_{\rm eff}(\varphi)$ (\ref{eq:poteff}) turns from mono-stable to multi-stable. \\
(iii) With even further increasing $\lambda>\lambda_2$ (also Sec.~\ref{sec:secondBifurc}), the system displays strong sensitivity to initial conditions (multiple steady-state orbits) and eventually irregular phase space patterns. When this domain is approached, the dc-current through the JJ  grows substantially in parallel with larger current variations depending on the driving amplitude. \\
(iv) For extremely strong driving $\lambda>\lambda_3\gg 1$ and finite friction (Sec.~\ref{sec:elevator}), the dynamics turns into a regular motion again with a large time scale separation between the motion happening within one of the local wells in $V_{\rm eff}(t,\varphi)$ and global dynamics exploring $V_{\rm eff}$ over a wide range of $\varphi$.

\subsection{Weak driving regime \label{sec:firstBifurc}}
As argued above, phase-space orbits in the weak driving regime remain basically ellipses, while the amplitude becomes nonlinear until a bifurcation occurs. Thus, the steady-state orbit (\ref{eq:ansatz}) is dominated by Fourier coefficients $\varphi_0, \varphi_{\pm 1}$. The corresponding equations are easily obtained by projecting Eq. (\ref{eq:BewglmitAnsatz})  on the respective Fourier modes as
\begin{widetext}
\begin{subequations} \label{eq:weakdrive}
\begin{equation}
\varphi_0+\lambda J_1(\varphi_1)\cos(\varphi_0-\theta_1)=0  \label{eq:P01}
\end{equation}
and with $\Lambda_1=\Omega^2-1$
\begin{eqnarray}
\varphi_1[ \cos(\theta_1)\Lambda_1+\gamma\Omega\sin(\theta_1)]-\lambda[J_0(\varphi_1)\sin(\varphi_0)-J_2(\varphi_1)\sin(\varphi_0-2\theta_1)] &=0 \label{eq:P11Om1a}\\
\varphi_1[- \sin(\theta_1)\Lambda_1+\gamma\Omega\cos(\theta_1)]-\lambda[J_0(\varphi_1)\cos(\varphi_0)+J_2(\varphi_1)\cos(\varphi_0-2\theta_1)] &=0 \label{eq:P11Om1b}\,.
\end{eqnarray}
\end{subequations}
\end{widetext}
This set of equations for $\{ \varphi_0, \varphi_1,\theta_1\}$ can be solved approximately for  small $\varphi_0, \varphi_1$ by exploiting that for the Bessel functions  one has $J_k(x)\sim O(x^k)$ for $|x|\ll 1$. One first gains from (\ref{eq:P01}) in leading order $\varphi_0\approx -\lambda J_1(\varphi_1)\cos(\theta_1)\approx -\lambda\varphi_1 \cos(\theta_1)$ so that driving dependent terms in (\ref{eq:P11Om1a}) are of higher order in the small parameter $\lambda$.
This yields the phase of oscillations,
\begin{equation}
\tan(\theta_1)= -\frac{\Lambda_1}{\Omega\gamma}\,
\end{equation}
 with $\theta_1=0$ at resonance. Inserting this result into (\ref{eq:P11Om1b}) leads close to resonance and in leading order in $\varphi_0$ to
\begin{equation}\label{eq:q1}
\varphi_1=\frac{\lambda}{\sqrt{\Lambda_1^2+\Omega^2\gamma^2}}\, [J_0(\varphi_1)+J_2(\varphi_1)]\, .
\end{equation}
 The known result for a driven harmonic oscillator is regained for $|\varphi_1|\ll 1$ while for somewhat larger driving nonlinearities in the Bessel functions tend to play a role. This type of orbit, with amplitude named $\varphi_{1}^I$ henceforth, which continually evolves from the harmonic-oscillator type of solution, is the only stable orbit until at a critical driving strength $\lambda_1$ a second solution appears with amplitude $\varphi_{1}^{II}$.
 \begin{figure}[tbh]
	\includegraphics[width=8.6cm]{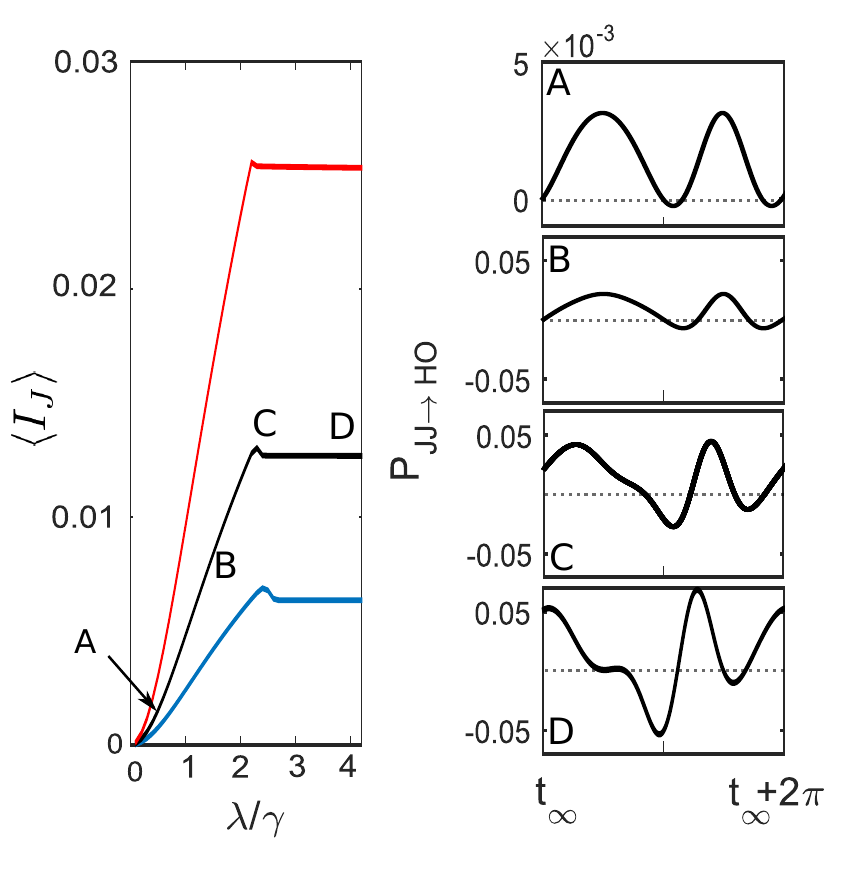}
	\caption{(color online) Left: Mean dc-current $\langle I_J\rangle$ through the JJ in steady state and on resonance $\Omega =1$ as a function of the scaled driving amplitude $\lambda/\gamma$ for (from bottom to top) $\gamma$=0.005 (blue), 0.01 (black, cf. Fig.~\ref{fig:OverviewSec3}), 0.02 (red).\\
	Right: Power transfer $P_{{\rm JJ}\to {\rm HO}}$ from the JJ to the resonator during one period of the oscillations in the steady state at $t_\infty=n\cdot 2\pi$,$n\gg 1$, $n\in \mathbb{N} $. The power transfer is shown for $\gamma=0.01$ and at four different driving strengths (also marked in the left panel): $\lambda/\gamma$=0.1 (A), 2 (B), 3 {(}C) and 3.5 (D). While the harmonic-oscillator type-I solutions gain energy from the drive nearly during the whole oscillation cycle (A and B), for type-II oscillations the energy gained during part of the cycle is partly flowing back to the drive in other parts (C and D). In consequence, increasing the driving strength beyond $\lambda_1/\gamma\approx 2.9$ does not further increase oscillation amplitude and current.	}
	\label{fig:EnergieuebertragfirstBifurc}
\end{figure}

In contrast to the harmonic-oscillator type orbit $\varphi_{1}^{I}$, the new orbit $\varphi_{1}^{II}$ exists even in absence of dissipation, and it is in this limit that it can easily be found analytically: putting $\gamma=0$ in  (\ref{eq:P11Om1a}), (\ref{eq:P11Om1b})  taken at  resonance $\Lambda_1=0$, we assume $\varphi_0=0$ to find the phase $\theta_{1}^{II}=\pi/2$ from \eqref{eq:P01}. Its amplitude then follows via \eqref{eq:P11Om1b} from
\begin{equation} \label{eq:amplphi12}
J_0(\varphi_1)-J_2(\varphi_1) = 2\frac{d}{d\varphi_1}J_1(\varphi_1)=0
\end{equation}
as $\varphi_{1, II}\simeq 1.841$ independent of the driving amplitude.
For finite friction and away from resonance orbits are obtained numerically from (\ref{eq:Langevindimless}).

It turns out that in a range beyond the threshold $\lambda_1$ the type-II orbit is the only stable one, while the harmonic-oscillator type-I orbits become unstable at this bifurcation point. The threshold  $\lambda_1$ is determined by the condition that the amplitudes of both solutions match, $\varphi_1^I(\lambda_1)=\varphi_1^{II}$, i.e., according to (\ref{eq:q1}) $\lambda_1/\gamma=\varphi_{1}^{II}/[2J_0(\varphi_{1}^{II})]\approx 2.912$.
In phase space both types of solutions display ellipsoidal orbits.
The transition from type-I to type-II solutions at the bifurcation point $\lambda_1$ has also been found in the classical limit of a quantum description within rotating-wave approximation in \cite{Armour13, Gramich13}.

An intuitive understanding of the saturation of the oscillation amplitude, when the driving strength is increased beyond $\lambda_1$ and, indeed, of the nature of the type-II orbit is offered by the numerical results in Fig.\ \ref{fig:EnergieuebertragfirstBifurc}.
How the resonator acts back onto the charge transfer in the JJ below/above the threshold is monitored by the energy transferred from the JJ to the resonator [cf.~(\ref{eq:dissEnergy})], i.e., the power $P_{{\rm JJ}\rightarrow {\rm HO}}(t)=-\lambda\dot{\varphi}(t)\sin[\varphi(t)+\Omega t]$ in the right panel of Fig.~\ref{fig:EnergieuebertragfirstBifurc}. Sufficiently below the threshold $\lambda_1/\gamma\approx 2.9$, energy is nearly unidirectionally injected into the resonator, i.e., the drive pushes energy into the oscillator at each time of the oscillation cycle. The phase shift of the type-II oscillations, however, results in an oscillation which extracts energy from the drive during one part of the oscillation cycle, and pushes energy back during another part. Increasing the driving strength beyond $\lambda_1$ will thus increase energy in- and back-flow, but will not further increase the net gain over a full cycle; hence, the saturation of the oscillation amplitude at $\varphi_{1, II}\simeq 1.841$.
Experimentally, the transition from type-I to type-II orbits is seen as a saturation in the dc-current through the JJ, i.e., $\langle I_J\rangle=\lambda \langle\sin[\varphi(t)+\Omega t]\rangle_\Omega$ , where the time average $\langle\cdot\rangle_\Omega$ is taken in steady state and over several oscillation periods, see left panel of Fig.~\ref{fig:EnergieuebertragfirstBifurc}. The dc-­current resulting from this averaging is (approximately) proportional $\langle\varphi_1^2\rangle$ [analytically found from Eq.~(\ref{eq:q1}) and (\ref{eq:amplphi12})]. Physically, this proportionality origins from balancing dissipation (proportional to the stored energy) and power input. Accordingly, the quadratic dependence $ \langle I_J\rangle\propto \lambda^2\propto E_J^2$ in the regime of very weak driving passes over to a linear dependence for somewhat stronger driving but still before the threshold. This changeover is reminiscent of the changeover from the perturbative domain of sequential (Coulomb blockade) to the regime of coherent charge transfer (phase coherent regime). Note that the various numerically found current curves in Fig.~\ref{fig:EnergieuebertragfirstBifurc} approximately just scale with $\gamma$, but for stronger damping the bifurcation threshold is shifted below the result $\lambda_1/\gamma\approx 2.9$ calculated for the $\gamma\rightarrow 0$ limit above.

\subsection{Beyond the first threshold\label{sec:secondBifurc}}
\begin{figure}[t]
	\begin{center}
	\includegraphics[width=8.6cm]{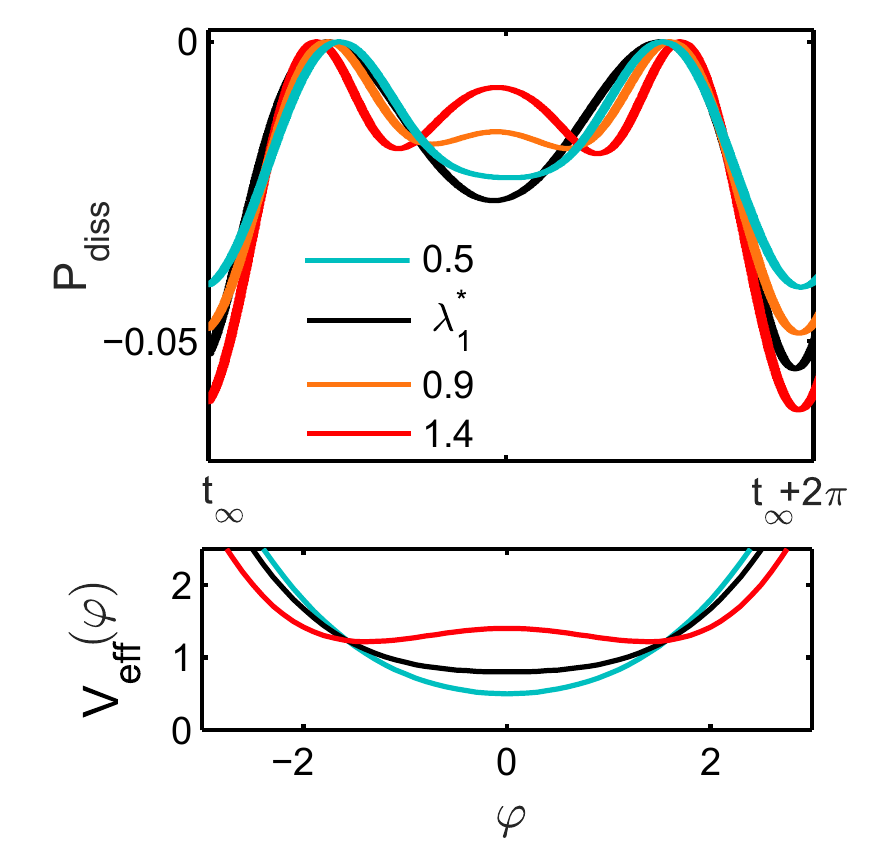}
	\caption{(color online) Top: Dissipated power $P_{\rm diss}$ from the driven JJ+resonator system for $\gamma=0.01$, $\Omega=1$ and various driving strengths $\lambda_1<\lambda<\lambda_2$ around the second bifurcation $\lambda_1^*$. Bottom: Snap-shots of the effective time-dependent potential $V_{\rm eff}(\varphi)$ (\ref{eq:poteff}) for various driving strengths (see top for the color code) beyond the first threshold $\lambda>\lambda_1$ and at times $\Omega t=(2n+1)\pi$, $n \in \mathbb{N}$.}
	\label{fig:Edisslam2}
	\end{center}
\end{figure}
\begin{figure*}[t]
	\begin{center}
	\includegraphics[width=0.95\textwidth]{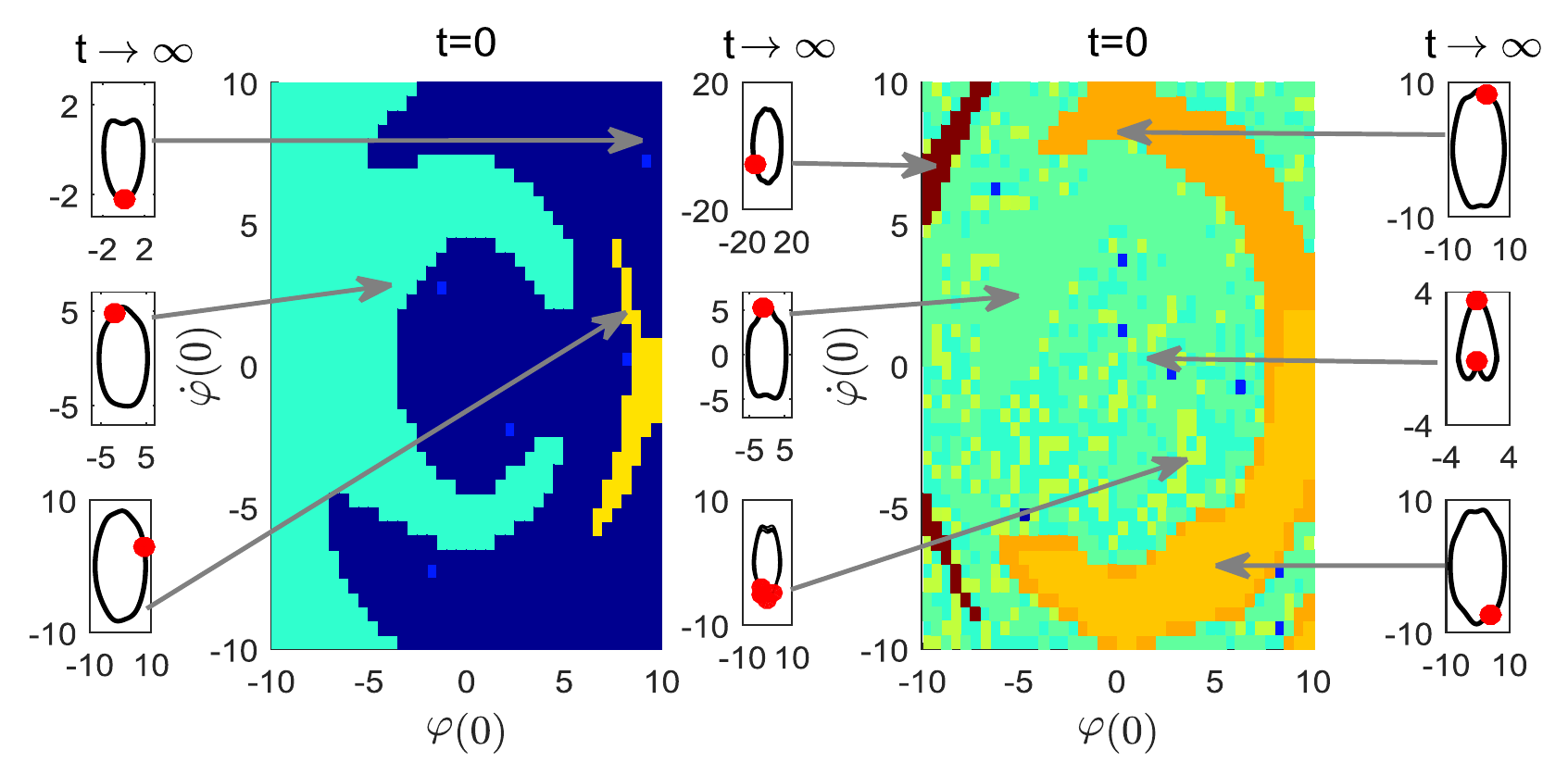}
	\caption{(color online) Various classes of steady-state orbits (outer panels) depending on the initial conditions (inner panels) for $\gamma=0.01$ and strong $\lambda=1$ (left) and very strong driving $\lambda=2.455$ (right). The color code for the initial conditions in phase space corresponds to a particular type of steady-state orbit in phase space. Red dots result from Poincar\'e plots and indicate  points of return for steady-state orbits after one period $2\pi/\Omega$.}
	\label{fig:Monster0}
	\end{center}
\end{figure*}

Beyond the first threshold $\lambda>\lambda_1$, type-II orbits dictate the dynamics until a second bifurcation occurs at  $\lambda_1^*\approx 0.8$ with only a very weak dependence on the friction strength in this underdamped regime.  The emergence of this class of orbits is characterized by a substantial deformation of the ellipsoidal phase-space structure (see Fig.~\ref{fig:OverviewSec3}), which is related to a changeover of the effective potential $V_{\rm eff}(\varphi)$ (\ref{eq:poteff}) from being essentially mono-stable to dominantly multi-stable (see Fig.~\ref{fig:Edisslam2}).

A particle moving (rather weakly damped) in the time-dependent effective potential will encounter a deep well during its passage through $\varphi=0$ in one direction, while on the way back in the second part of its oscillation cycle (see lower panel of Fig.~\ref{fig:Edisslam2}), the well is less pronounced (corresponding to the situation below $\lambda_1^*$) or  even turns into a barrier around $\varphi=0$ (situation well above $\lambda_1^*$). Indeed, in the  power dissipated from the driven system into the reservoir $P_{\rm diss}(t)=-\gamma \dot{\varphi}^2(t)$ (upper panel of Fig.~\ref{fig:Edisslam2}), the maximal amplitude at  $\Omega t= 2 n \pi\;(n \textrm{ integer})$ shows a local maximum right around the bifurcation value $\lambda_1^*$, whereas the speed of the fictitious particle (and concurrently $P_{\rm diss}$) at $\Omega t= (2 n+1) \pi$ is more and more reduced and even develops a local minimum when passing through $\varphi=0$.
While this second bifurcation has almost no effect on the dc-current (cf.~Fig.~\ref{fig:OverviewSec3}), its appearance is detectable in the discussed features of the dissipated power.

With further increasing the driving $\lambda>\lambda_1^*$ and initial conditions close to the phase space origin, further bifurcations occur that we do not need to discuss in detail here. It is important to note though that each bifurcation is associated with a change in stability meaning that only the newly emerging orbits determine the dynamics beyond each bifurcation threshold. Independent of the appearance  of new orbits, the dc-current $I_{\rm J, dc}$ stays basically constant over a wide range of driving amplitudes $\lambda_1<\lambda<\lambda_2\approx1.6$ (see Fig.~\ref{fig:OverviewSec3}). Similar to $\lambda_1^*$ the numerical value of $\lambda_2$ depends only very weakly on the friction strength in the underdamped regime.

For the driving strengths considered so far, it has been sufficient to consider initial conditions $\{\varphi(0), \dot{\varphi}(0)\}$ close to the phase space origin only.
Now, that steady-state orbits tend to explore larger domains in phase space, one may wonder about the impact of initial conditions located in these regions. This is illustrated in Fig.~\ref{fig:Monster0} (left), where classes of steady-state orbits are studied depending on their initial conditions for moderately strong driving $\lambda=1$, i.e., $\lambda_1<\lambda<\lambda_2$.

Indeed, for different sets of initial conditions $\{\varphi(0), \dot{\varphi}(0)\}$, one now asymptotically finds three different classes of steady-state orbits in phase space.
This dynamical multi-stability reflects the multi-stability of the effective potential $V_{\rm eff}(\varphi)$ due to a strong nonlinearity. However, in this regime of driving-strengths, $\lambda_1<\lambda<\lambda_2$, the existence of multiple types of steady-state orbits will usually not be of experimental relevance. This is due to the fact, that multi-stability only occurs if the fictitious particle leaves local wells in  $V_{\rm eff}$ which requires at least an energy of $2\lambda$ .
This could, in principle, be done by preparing the system initially (a difficult task though) such that $E_{\rm HO}(\dot{\varphi}(0),\varphi(0))>2\lambda$: See the dark blue area in the left panel of Fig.~\ref{fig:Monster0}, indicating initial conditions around the phase space center which lead to one and the same steady-state orbit.
Starting from an equilibrated circuit, however, domains of initial conditions leading to additional steady-state orbits, do play a role only at highly elevated temperatures, $k_B T/m\omega_0^2>2\lambda$.
For the current experimental situation, this regime is not relevant.

 For $\lambda>\lambda_2\approx1.6$, one enters again a qualitatively new regime. It is characterized by a sharp rise of the dc-current through the JJ  by almost an order of magnitude (see Fig.~\ref{fig:OverviewSec3}). Now, even for initial conditions close to the phase space origin, multiple types of stable steady-state orbits exist, both exploring large domains in phase space, see Fig.~\ref{fig:Monster0} (right). Some of these  orbits are covered only after multiples of the fundamental period $2\pi/\Omega$, in contrast to the regime $\lambda<\lambda_2$ [cf.~single (multiple or single) Poincar\'e-plot points in the orbits of the left (right) panel of Fig.~\ref{fig:Monster0}].
 Hence, the sensitivity with respect to initial conditions grows substantially, thus marking the onset of irregular, chaotic-like behavior for sufficiently large driving amplitudes  $\lambda>\lambda_2$.  A detailed analysis of properties and characteristics of possible chaotic dynamics in this domain is beyond the scope of this paper and will be presented elsewhere.

\subsection{From multi-well to elevator dynamics \label{sec:elevator}}
\begin{figure}[b]
	\begin{center}
	\includegraphics[width=8.6cm]{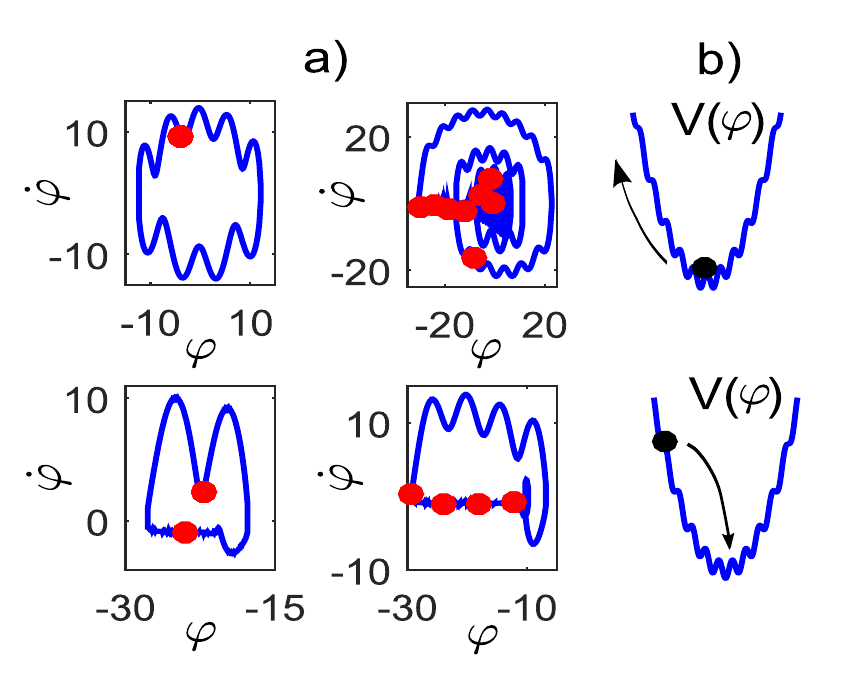}
	\caption{(color online) (a) Phase space portraits of steady-state orbits with Poincar\'e points (red) in the regime of extremely strong driving with $\lambda=30\gtrsim \lambda_3$ and $ \gamma=0.01$ (top left), 0.025 (top right), 2.5 (bottom right), 5 (bottom left). Sketched in (b) is the particle dynamics in  the effective potential $V_{\rm eff}(\varphi)$, see Eq.~(\ref{eq:poteff}), corresponding to phase space orbits in the right panels of (a).  For moderate damping the particle is trapped in one of the local wells, `elevated' upwards in $V_{\rm eff}(\varphi)$ [the motion towards large negative $\varphi$-values with small $\dot{\varphi}$ in (a)] until its escape and consequent retrapping.
	\label{fig:elevator}
	}
	\end{center}
\end{figure}

Interestingly, in the regime of extremely strong driving, $\lambda>\lambda_3\gg 1$, (and for finite friction) regular dynamics dominates again (see Fig.~\ref{fig:elevator}). At this driving strength the periodic part of $V_\mathrm{eff}$ is pronounced enough to create a multitude of local minima in the superimposed quadratic potential, see Fig.~\ref{fig:elevator}(b).
We mention in passing that this multi-well pattern of $V_\mathrm{eff}$ has some analogy to the potential profile of superconducting quantum interference devices (SQUIDs). In contrast to SQUIDs, however, the potential $V_{\rm eff}(\varphi)$, cf. Eq.\ \eqref{eq:poteff}, combines a static quadratic with a {\em time-dependent} sinusoidal potential.

The dynamics is then easily understood; most simply in the completely underdamped and the strongly overdamped cases.
In the completely underdamped case [upper left of Fig.~\ref{fig:elevator}(a)] the particle essentially undergoes a simple oscillation in the quadratic potential over a wide $\varphi$ region, running with high energy over the potential wells, which then causes slight wiggles in the phase-space orbit.

For somewhat stronger friction [upper right of Fig.~\ref{fig:elevator}(a)], a fictitious particle runs periodically through a cycle of localized and de-localized motion: Starting somewhere in the low energy sector, it gets trapped quickly in one of the local minima of the potential close to the global minimum of $V_{\rm eff}(\varphi)$ (\ref{eq:poteff}). It is then transferred up in energy by the driving term $\propto \cos(\varphi+\Omega t)$ while being trapped in this local well. During this process the potential barrier of the respective local well shrinks until the particle can escape to run towards the global minimum while loosing substantial energy so that it gets trapped close to the global energy minimum again. This type of ''elevator'' dynamics leads to increasingly simpler phase space patterns towards the overdamped regime. Physically, during the trapping period $|\varphi|$ grows almost linearly with $|\dot{\varphi}|\sim 1/\Omega$, whereas the motion towards the global minimum is associated with an almost instantaneous drop in amplitude accompanied by a large increase in momentum. As a consequence, based on the second Josephson relation $V(t)\propto \dot{\varphi}(t)$, one expects to observe strong voltages pulses with frequencies much lower than the driving frequency $\Omega$.

\section{Higher order resonances \label{sec:Resonances} }

So far we have considered driving the system at (or close to) the eigenfrequency, $\omega_0 \equiv 1$, of the resonator. At sufficiently strong driving, the system's response then contains higher harmonics. It is, thus, also interesting to drive at frequencies $\Omega \not \approx 1$, where higher or subharmonics of the drive can become resonant with the eigenfrequency. Physically, the drive is detuned, of course, simply by changing the applied dc-voltage bias.

\begin{figure}[bt]
	\begin{center}
	\includegraphics[width=8.4cm]{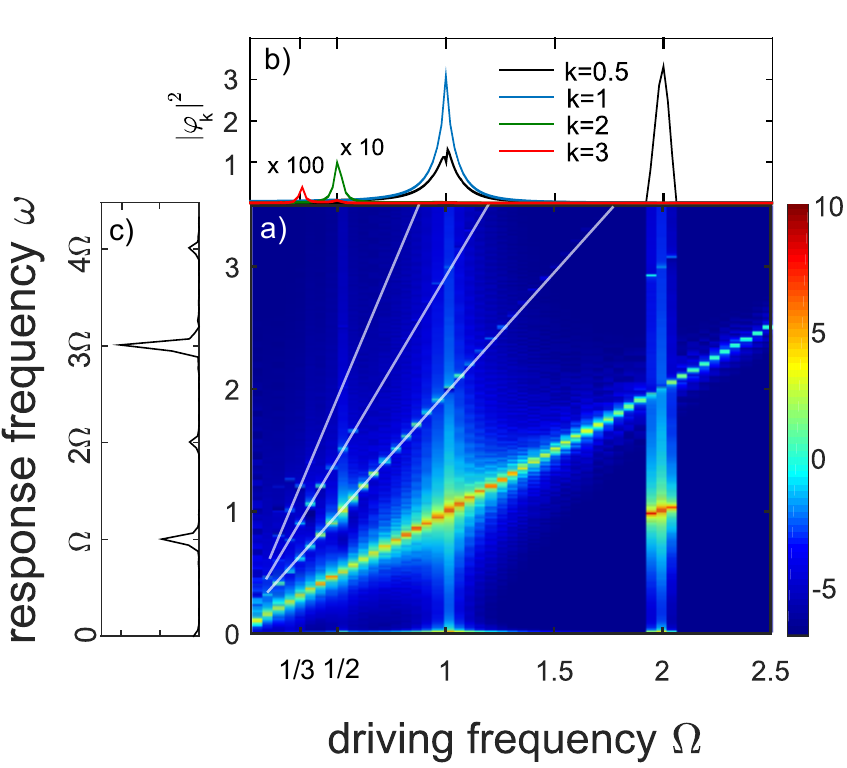}
		\caption{(color online) a) Power spectrum, $\ln(| \tilde{\varphi}(\omega)|^2)$, of steady-state orbits $\varphi(t)$ for a range of driving frequencies $\Omega$ in the underdamped regime $\gamma=0.1$ and at moderate driving amplitude $\lambda=0.2$.  For the sake of presentation the spectrum is numerically broadened by taking the Fourier transform of a finite time signal.
b) Dominant components $|\varphi_k|^2$ of $\varphi(t)$ [see (\ref{eq:ansatz})], corresponding to lines $\omega=k\Omega$ in a). Panel c) shows a cut of the power spectrum at the (shifted) subharmonic resonance $\Omega =0.365\approx \Omega_{\frac{1}{3}}$ at a larger driving amplitude $\lambda=0.7$, where the system dominantly responds with $\omega=3\Omega\approx 1$.
	\label{fig:Spektrum}
	}
	\end{center}
\end{figure}

\begin{figure}[t]
	\begin{center}
	\includegraphics[width=7cm]{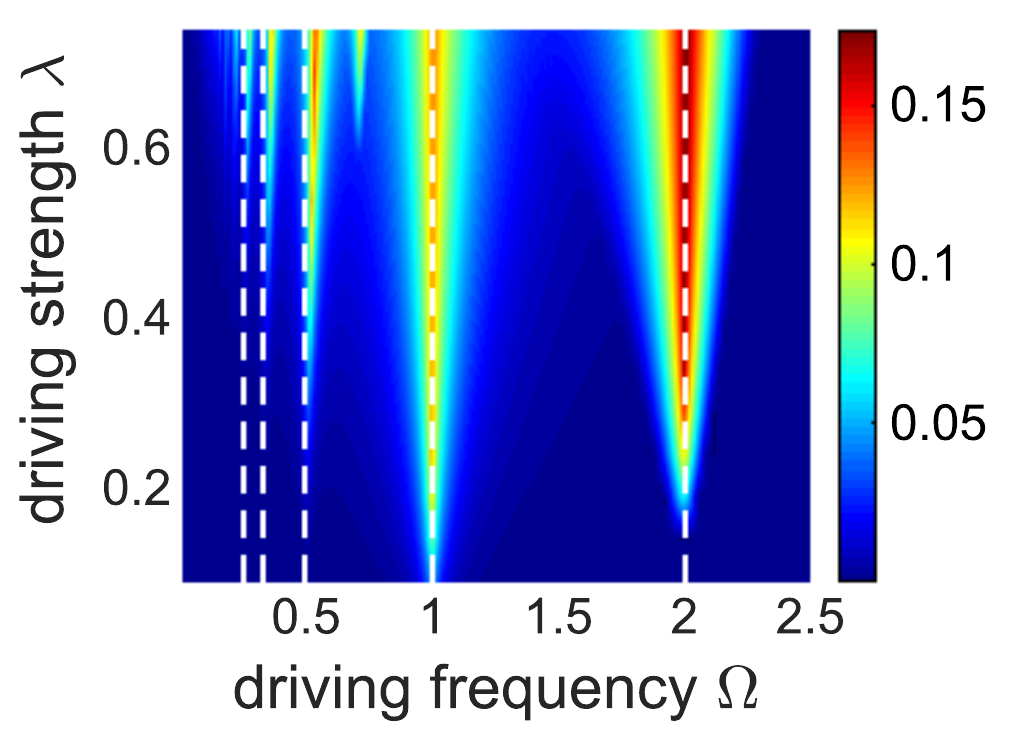}
	\caption{(color online) DC-current through the JJ vs.\ driving frequency $\Omega$ and driving amplitude $\lambda$ for $\gamma=0.1$. For weak driving the system is nearly linear and dominated by the fundamental resonance, $\Omega=\omega_0=1$, while for stronger driving the subharmonic resonances at $\Omega=\Omega_{\frac{1}{n}}=\omega_0/n$ and the multi-photon resonances at $\Omega= \Omega_n=n \omega_0$ become apparent. Note, the sharp onset of two-photon processes at $\lambda_c \approx \gamma \Omega=0.2$ described in Subsec.~\ref{sec:21}, while subharmonic resonances increase smoothly and shift with increasing $\lambda$ as discussed in Subsec.~\ref{sec:subharmonics} (cf. also Fig.~\ref{fig:Resonanzkurve(21)0k1}).
	}
	\label{fig:currentOverview}
	\end{center}
\end{figure}

Figure~\ref{fig:Spektrum} shows the steady-state power spectrum, $| \tilde{\varphi}(\omega)|^2$, i.e., the response of the system at frequency $\omega$ if driven at $\Omega$. Such spectra have been recently investigated experimentally in Ref.~[\citen{Chen14}]. Resonances [i.e., stable steady-state solutions of (\ref{eq:Langevindimless}) or (\ref{eq:BewglmitAnsatz})  with large amplitudes] are found
for driving frequencies $\Omega_n=n$ for integer  $n\neq 0$,  where the system responds at $\omega=1, 2, 3,\ldots$, and for driving frequencies $\Omega_{\frac{1}{n}}=1/n$ with response frequencies $\omega=1/n, 2/n, 3/n, \ldots$. 
The situation, where the system is driven with $\Omega$ and responds with $\omega<\Omega$ is called down-conversion, the situation, where it responds with $\omega>\Omega$ up-conversion.  
Apparently, for a driving frequency $\Omega_2$ the system dominantly responds with $\omega=1$ while contributions of higher harmonics are weak. In contrast, driving with $\Omega_{\frac{1}{2}}$ one observes a response in which dominantly frequencies with  $\omega=1/2$ but also the first few higher harmonics $\omega= 1,\,3/2, \ldots$ are present.
While the power spectrum, $| \tilde{\varphi}(\omega)|^2$, shown here gives a direct intuitive link to the system dynamics, comparable information can also be extracted from the directly measurable spectrum of light emission.
Before we turn to details of its resonant features below, let us address the consequences of varying the dc-bias (and thus the driving frequency) for the dc-current through the JJ.
As depicted in Fig.~\ref{fig:currentOverview}, current peaks appear indeed at driving frequencies $\Omega_n$ and $\Omega\approx \Omega_{\frac{1}{n}}$ with $n\geq 1$ integer with a characteristic shift towards  $\Omega>\Omega_{\frac{1}{n}}$ in the subharmonic domain.
 These features may remind of Shapiro steps \cite{Barone82} known from JJs subject to dc- and ac-voltages of the form $V(t)=v_{dc} + v_{ac} \cos(\Omega_{ac} t)$, an interpretation which for the present situation is somewhat confusing though: The conventional argument for the existence of Shapiro steps is that resonances appear whenever the energy gap induced by the dc-voltage $2 e v_{dc}$ matches multiples of ac-photon quanta $\hbar\Omega_{ac}$, i.e., $2 e v_{dc}/\hbar= n \Omega_{ac}$, $n\geq 1$ being integer. For the present situation, however, one could argue in two ways (we temporarily return to physical dimensions): (i) The dc-voltage which determines the driving frequency $\Omega=2 e V/\hbar$ must be multiples of the resonator frequency $\omega_0$ implying $\Omega=n \omega_0$, $n\geq 1$ integer, or (ii) the excitation of the resonator by one energy quantum  $\hbar\omega_0$ requires multiples of ac-photon quanta $n\hbar\Omega$, $n\geq 1$ integer, i.e., $\Omega=\omega_0/n$. Along these lines, one could interprete either the subharmonic $\Omega_{\frac{1}{n}}$ or the higher harmonic $\Omega_n$ resonances as Shapiro steps.
The fundamental difference to the conventional set-up to observe Shapiro steps is that there the dynamics of the Josephson phase $\phi$ is fixed by the external voltage according to $ 2 e V(t)=\dot{\phi}/\hbar$, while here the Josephson phase appears as a {\em dynamical} degree of freedom fixed by the dynamics of the resonator phase $\varphi$ [see Eq.~(\ref{eq:eom1})].

\subsection{Integer multi-photon processes \label{sec:21}}
Now, we start with multi-photon resonances $\Omega_n=n$, where the system asymptotically responds mostly with frequency $\omega=1$ (down-conversion).  In particular, we focus on the generic case $\Omega=2$ (see  Fig.~\ref{fig:Spektrum}) which can be interpreted as a parametric resonance.

The general argument for the appearance of this type of resonances can be directly read off from the equation for the Fourier mode amplitudes (\ref{eq:BewglmitAnsatz}): For $\Omega=n$ we seek orbits with a time dependence dominated by $\cos(t)$ implying that Fourier coefficients $\varphi_{\frac{1}{n}}$ with $k n=1$ dominate the expansion (\ref{eq:ansatz}). Accordingly, in (\ref{eq:BewglmitAnsatz}) $F_\pm \approx F_{\frac{1}{n}}^\pm(\varphi_{\frac{1}{n}})$ and upon projecting onto orbits with $ {\rm e}^{\pm i t}$, one arrives at time-independent equations for the coefficients $\varphi_{\frac{1}{n}}$ which include Bessel functions $J_{n-1}(\varphi_{\frac{1}{n}})$ and $J_{n+1}(\varphi_{\frac{1}{n}})$ [see (\ref{eq:fpmbessel})].

In the particular case of $\Omega=2$, we may write for small amplitudes $\lambda \sin[\varphi(t)+2 t]\approx \lambda[\sin(2t)+\varphi(t) \cos(2t)]$, thus giving in the equation of motion (\ref{eq:Langevindimless}) effectively rise to a parametrically driven harmonic oscillator scenario.
For this system steady-state solutions only exist below a critical, damping dependent value $\lambda_c = 2\gamma$, while above that threshold oscillation amplitudes will grow infinitely in time. Of course, here the nonlinearity of the full problem prevents this divergence to occur. Then, the simplest ansatz, $\varphi(t)\approx  \varphi_{\frac{1}{2}} \cos(t+\theta_{\frac{1}{2}})$, taking only oscillations at the oscillator's resonance frequency into account, yields for the amplitudes
 \begin{equation}
[J_1(\varphi_{\frac{1}{2}})+J_3(\varphi_{\frac{1}{2}})]/\varphi_\frac{1}{2}=\gamma/\lambda\, .
\end{equation}
Solutions of this equation only exist above the parametric-resonance threshold,  $\lambda > \lambda_c = 2\gamma$, with a phase $\theta_\frac{1}{2} \approx - \pi/4$ (cf. also [\citen{Armour13}]).
The full solution with an additional  contribution $|\varphi_1|\ll |\varphi_{\frac{1}{2}}|$ and $\theta_1\approx \pi/2$ shows similar threshold behavior with some quantitative corrections (cf. the threshold in Fig.~\ref{fig:currentOverview} and  Fig.~\ref{fig:Resonanzkurve(21)0k1}).

\subsection{Subharmonic resonances\label{sec:subharmonics}}
In the driving-frequency range $\Omega<1$ below the fundamental resonance, the system will dominantly respond with a few higher harmonics $\omega=\Omega,\,2\Omega,\dots$, see Fig.~\ref{fig:Spektrum}. Resonances appear approximately at driving frequencies $\Omega_{\frac{1}{n}}$, where one of these harmonics matches the eigenfrequency $\omega_0=1$. This resonant response results, in particular, in a strong contribution to the dc-current close to $\Omega_{\frac{1}{n}}$, where a shift of the resonance with increasing driving strength can be observed, see Fig.~\ref{fig:currentOverview}.

As a specific example, we consider the shift of the resonance close to $\Omega=1/2$ for $\gamma\to 0$. Taking then an ansatz for the steady-state orbits (\ref{eq:ansatz})  of the form
$\varphi(t)\sim \varphi_1 \cos(\Omega t+\theta_1)+\varphi_{2} \cos(2\Omega t+\theta_2)$
with $\Omega \approx \Omega_\frac{1}{2}$
and considering (\ref{eq:BewglmitAnsatz}) we find, that the Fourier coefficients for this type of steady-state orbits have to be gained from nonlinear equations including products of Bessel functions $J_0(\varphi_2) J_1(\varphi_1)$, $J_2(\varphi_2) J_1(\varphi_1)$.
Assuming for sufficiently strong driving a dominant response with $2 \Omega \approx 1$ and thus $\varphi_1\ll 1, \,\varphi_2$, we take into account only the lowest order of the Bessel functions for $\varphi_1$ and find the resonance condition
$
\Lambda_2 = -\lambda \cos{\theta_1} \varphi_1 J_1(\varphi_2)/\varphi_2 >0
$
from the $2 \Omega$-projection of Eq.\ \eqref{eq:BewglmitAnsatz}. That this shift is indeed positive, as seen in Fig.\ \ref{fig:currentOverview}, can be analytically confirmed for this limit by analyzing the $\Omega$-projection. Increasing the driving strength, the shift that depends on the mixing between the driving frequency and the first higher harmonics grows further. Remarkably, at least for moderate driving similar shifts do not occur for the $\Omega_n$-resonances (including the fundamental one), as apparent from the numerical data in Fig.~\ref{fig:currentOverview} and analytically following  Eq.\ \eqref{eq:weakdrive} and Sec.\ \ref{sec:21}.

 Note, that the multi-photon processes and subharmonic resonances discussed here and shown in Figs.\ \ref{fig:Spektrum}, \ref{fig:currentOverview} all occur for comparatively weak driving; for strong driving similarly rich behavior for driving at sub- and higher-harmonic resonance frequencies may be expected,  as found in Sec.\  \ref{sec:secondBifurc}, \ref{sec:elevator}  for the fundamental resonance.

\subsection{Energy transfer}
As already discussed for the fundamental resonance in Sec.~\ref{sec:firstBifurc}, the energy transfer from the driving source to the resonator is another tool to reveal details of the nonlinear dynamics. Experimentally, it is accessible as (mean) photon emission from the resonator.
\begin{figure}[bt!]
	\begin{center}
	\includegraphics[width=8cm]{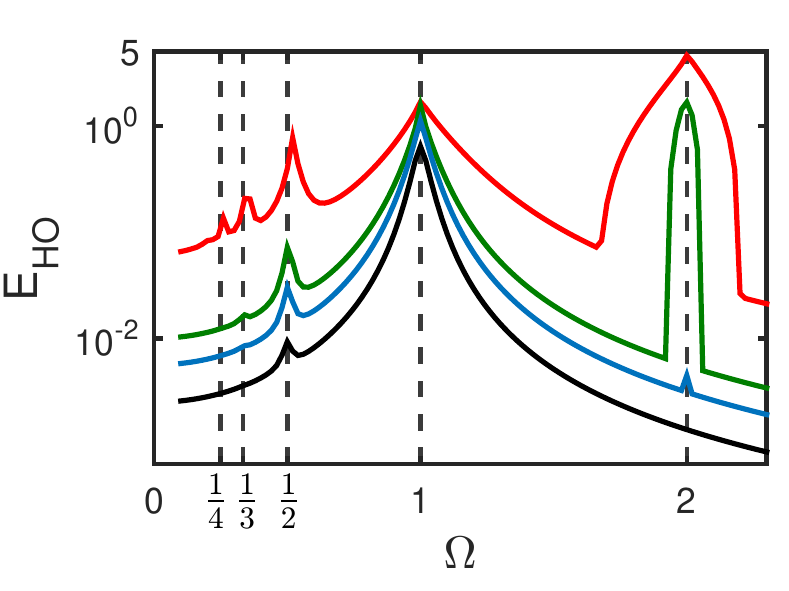}
	\caption{(color online) Time-averaged steady-state energy in the resonator for $\gamma=0.1$ and at different driving strengths: $\lambda=$0.1 (black), 0.15 (blue), 0.2 (green), 0.5 (red).	For the two-photon resonance, $\Omega=2$, there is a rather sharp threshold at $\lambda \gtrsim\lambda_c=2\gamma$, while subharmonic resonances, $\Omega\approx1/n$  gradually increase and shift with increased driving strength.
	\label{fig:Resonanzkurve(21)0k1}
	}
	\end{center}
\end{figure}
Here, we depict resonance curves of the time averaged resonator energy $E_{\rm HO}=\langle\dot{\varphi}^2/2+\varphi^2/2\rangle_\Omega$, see Fig.~\ref{fig:Resonanzkurve(21)0k1}. As expected, the resonances discussed in the previous sections appear in form of pronounced peaks at frequencies $\Omega_n$ and $\Omega_{\frac{1}{n}}$. In the subharmonic regime the shift in the resonances towards $\Omega>\Omega_{\frac{1}{n}}$ for larger $\lambda$ is seen as well. We note also the threshold $\lambda>\lambda_c$ for the occurrence of the parametric resonance at $\Omega_2$.

\section{Thermal noise \label{sec:Thermal}}
In actual experimental realizations, noise stemming from various sources is always present and may sensitively influence the dynamics of the JJ+resonator device. In the Langevin equation (\ref{eq:Langevindimless}) we restricted ourselves to thermal noise related to the finite photon lifetime in the resonator via the fluctuation-dissipation theorem.  Another major source of noise may be local voltage fluctuations at the JJ that may induce charge localization (Coulomb blockade). However, for the present situation the impact of these fluctuations is of minor relevance.
Although assuming a purely classical regime implies, that temperatures are \emph{high} as compared to quantum fluctuations, the circuit is nonetheless operated at  temperatures sufficiently \emph{low} compared to other energy scales of the system, i.e., in dimensional units $k_{\rm B} T< m\omega_0^2, E_J$.
\begin{figure}[t]
	\begin{center}
	\includegraphics[width=8.6cm]{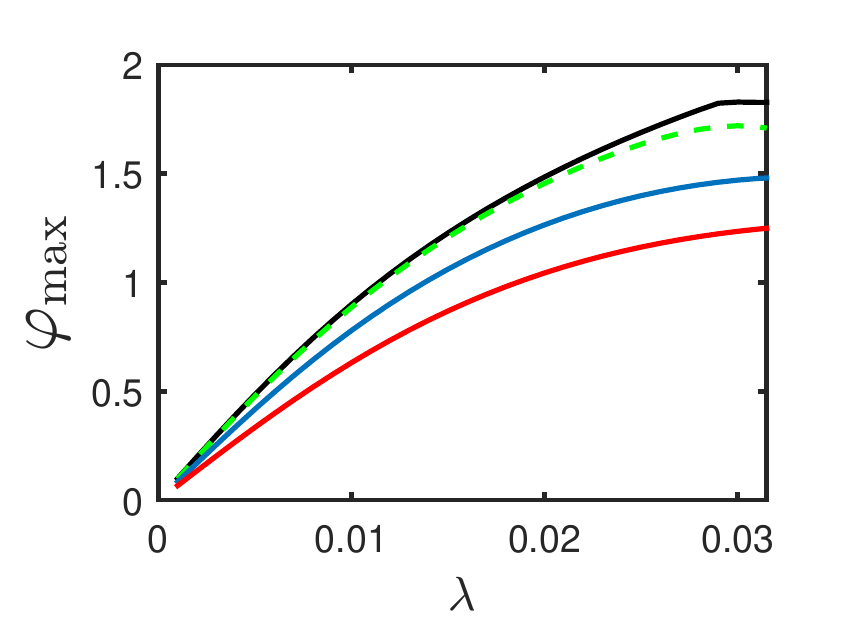}
	\caption{(color online) Mean steady-state oscillation amplitude, $\varphi_\mathrm{max}=\mathrm{max}\left[ \langle \varphi(t\rightarrow\infty)\rangle_\xi\right]$, vs.\ driving strength averaged over 10000 realizations of thermal noise at temperatures $k_{\rm B}T$=0 (black), 0.01 (green-dashed), 0.1 (blue), 0.25 (red) for friction parameter $\gamma=0.01$.}
	\label{fig:thermalBifurk}
	\end{center}
\end{figure}

The simplest
way to include finite temperature effects is to perform a thermal averaging over initial conditions, i.e., we start initially with a thermal distribution of phase space variables but then follow a deterministic time evolution according to (\ref{eq:Langevindimless}) for $\xi=0$.
 While somewhat inconsistent, this scenario accounts for the fact that precise initial conditions are experimentally not feasible. From a purely theoretical perspective, it allows to analyze the sensitivity of steady-state orbits onto initial conditions.

 The full description of thermal noise works with the Langevin equation (\ref{eq:Langevindimless}) and seeks for phase space orbits averaged over many noise realizations $\langle\varphi(t)\rangle_\xi, \langle\dot{\varphi}(t)\rangle_\xi$. In this latter situation,  asymptotically thermal fluctuations may induce transitions between
various steady-state orbits even when their respective initial conditions are well separated in phase space.

The main effect of an initial thermal distribution is apparent in the regime $\lambda>\lambda_2$ (see Fig.~\ref{fig:OverviewSec3}), where bare steady states tend to depend sensitively on the initial conditions. Accordingly, when averaged over an initial  thermal distribution, phase space structures in steady state are washed out. This in turn gives rise to relatively large current fluctuations $\langle (I_{\rm dc}-\langle I_{\rm dc}\rangle_0)^2\rangle_0$, where $\langle \cdots\rangle_0 $ denotes the average over initial conditions according to a thermal distribution. In fact, the size of current fluctuations directly indicates ranges in parameter space, where the underlying asymptotic dynamics displays either bifurcations or irregular behavior, cf.~Fig.~(\ref{fig:OverviewSec3}).

Thermal noise according to the full Langevin dynamics has a similar impact. Here, we focus on two domains, namely, the domain around the first bifurcation $\lambda\approx \lambda_1$ for $\Omega=1$ (see Sec.~\ref{sec:firstBifurc}) and the domain around the parametric resonance $\lambda\approx \lambda_c\approx 2 \gamma$ for $\Omega=2$ (see Sec.~\ref{sec:21}). 

For the first case,  in Fig.~\ref{fig:thermalBifurk} the mean steady-state amplitude  $\mathrm{max}\left[ \langle \varphi(t\rightarrow\infty)\rangle_\xi\right]$ is shown. Even in the weak driving regime the linear response of the system gets influenced by temperature as thermal fluctuations become larger and increasingly explore the nonlinearity of the effective potential. More substantial deviations occur for $\lambda\to \lambda_1$, where the bifurcation is increasingly smeared out and the overall oscillation amplitude decreases at elevated temperatures.

\begin{figure}[t]
	\includegraphics[width=8cm]{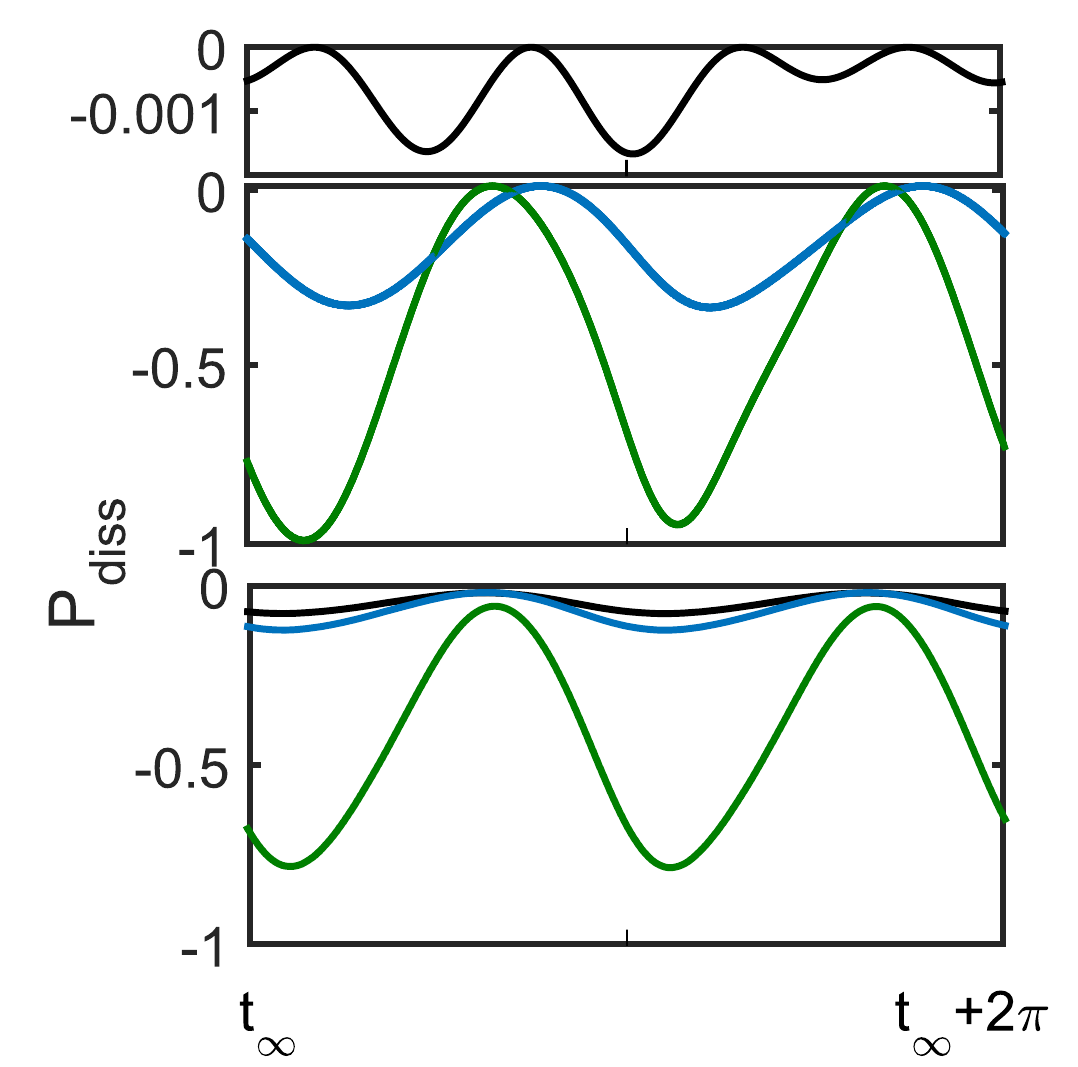}
	\caption{(color online) Mean dissipated power at temperature $k_{\rm B}T=0$ (top two panels) and $k_{\rm B}T=0.1$ (bottom) with $\gamma=0.1$ and $\Omega=2$ for various driving strengths $\lambda$=0.15 (black), 0.2 (blue), 0.5 (green). The threshold for the parametric  amplification in absence of noise is $\lambda_c\approx 2\gamma=0.2$.}
	\label{fig:thermalDownConversion}
\end{figure}
In the second case, the parametric-resonance threshold, the situation is a bit more intricate. Here, thermal noise may mix higher harmonics in the Fourier expansion (\ref{eq:ansatz}) with the consequence that down-conversion already occurs prior to the threshold $\lambda_c$, see Fig.~\ref{fig:thermalDownConversion}.
Accordingly, already for driving strengths below the zero-temperature threshold, thermal noise excites oscillations with the resonant frequency $\omega=\Omega_2/2=1$ with a drastically increased amplitude compared to the response at the driving frequency $\omega=\Omega_2$ in the absence of noise (e.g., an increase by about two orders of magnitude for the black lines in Fig.~\ref{fig:thermalDownConversion}). Above-threshold oscillations are somewhat reduced by temperature.
Nonetheless, a transition remains clearly visible, for example, in the mean dissipated power $\langle P_{\rm diss}(t)\rangle_\xi$. Furthermore, an offset emerges such that $\langle P_{\rm diss}(t)\rangle_\xi>0$ for all times, which can be related to the thermal energy continuously injected into the system.

\section{Conclusion and discussion\label{sec:Conclusion}}

In this paper we analyzed the classical dynamics of a circuit, where a single relevant resonator mode interacts with a dc-voltage biased JJ. This problem can be mapped onto the dissipative dynamics of a fictitious particle moving in a nonlinear, time dependent potential, where in contrast to conventional settings the nonlinearity appears as part of the driving, while the static part of the potential is purely harmonic. In the regime of moderate to large $Q$-factors (underdamped regime) and weak thermal noise, steady-state orbits and corresponding observables are determined by basically only two parameters, namely,  the dimensionless driving strength (Josephson coupling) $\lambda=E_J/m\omega_0^2$ and the dimensionless driving frequency $\Omega$ (in units of $\omega_0$).

At the fundamental resonance $\Omega=\omega_J/\omega_0=1$ this system displays a changeover from a linear response regime for weak driving towards a strongly nonlinear behavior for strong driving. The various dynamical domains leave their signatures in the dc-current flowing through the JJ and in the microwave power emitted from the resonator and are thus directly accessible experimentally.
Resonances are also found when driving with either higher harmonics ($\Omega=n$,  $n$ integer) or sub-harmonics ($\Omega\approx1/n$, $n$ integer), while the system responds with the fundamental frequency thus corresponding to processes of down- and up conversion.
These features can also be detected by either monitoring the dc-Josephson current or the radiated microwaves.
 Due to its high degree of tunability the resonator+JJ circuit thus allows to study the full wealth of classical nonlinear dynamics
 in one-dimensional driven, dissipative  systems. The impact of weak thermal noise is most prominent close to bifurcations of steady-state orbits.

At this point let us discuss a typical set of parameters for circuit designs that allows to access the physics discussed above:
The classical regime with weak thermal noise requires that $m\omega_0^2 \gg k_{\rm B} T\gg \hbar \omega_0$. For a resonance frequency of $\omega_0\sim 5\, $GHz, this can be realized with an $LC$-circuit with $C\sim 5\,$pF operated at temperatures $T\sim 150\, $mK. One then has $m\omega_0^2\sim 0.08\, $meV, $k_{\rm B}T\sim 0.015\, $meV, and $\hbar\omega_0\sim 0.003\, $meV. In the phase regime of the JJ, one further needs $E_J\gg E_C$ which applies for a typical $E_J\sim 1\, $meV. An external magnetic flux allows to tune this maximal Josephson coupling down to a factor of about 100, i.e., within the range \ $E_J\sim 0.01\ldots 1\,$meV. Present resonator designs have typical photon lifetimes over a wide range of $Q$-factors $Q\approx 10\ldots 10^4$ which coincides with the (strongly) underdamped regime.

Theoretically, at the fundamental resonance new dynamical domains are associated with driving parameters $\lambda_1, \lambda_1^*, \lambda_2 $, and $\lambda_3$ as depicted in Fig.~\ref{fig:OverviewSec3} which, given the above parameters, translates into the following coupling energies:
$E_{J, 1}\approx 0.02\, $meV, $E_{J, 1}^*\approx 0.05\, $meV, $E_{J, 2}\approx 0.1\, $meV, $E_{J, 3}(\lambda\approx 20)\approx 1.3\, $meV. Apparently, the first three coupling energies are easily accessible experimentally with only $E_{J, 3}$ lying at the edge of the range. The challenge here is that to cover the full range of driving amplitudes within one set-up from the linear response regime $E_J\ll E_{J, 1}$ to $E_{J, 3}$, requires to vary $E_J$ by a factor of somewhat more than 100.

Note that the superconducting gap $2\Delta$ defines an upper limit on the applied voltage $2eV<2\Delta$ which for Al junctions corresponds to $\omega_J\ll 500\, $GHz, thus allowing also for driving at higher harmonics $\Omega_n$.
Following this discussion, we are confident that the classical, nonlinear dynamics analyzed in this work is indeed accessible in realistic circuits similar to those in reference [\citen{Chen14}] and opens the door to study the interplay of driving, dissipation and resonances under well controlled conditions.

\begin{acknowledgments}
 This work benefitted from discussions with M. \mbox{Blencowe}, F. Portier, and A. Rimberg. JA and BK thank the Department of Physics and Astronomy, Darmouth College, Hanover, USA, for the kind hospitality.
Financial support from the Carl-Zeiss-Foundation (SM), the Harris-Foundation (JA), and the Deutsche Forschungsgemeinschaft through AN336/6-1 is gratefully acknowledged.

\end{acknowledgments}

\bibliography{Bibfile}

\end{document}